\documentclass[epj]{webofc}
\usepackage[varg]{txfonts}   
\wocname{EPJ Web of Conferences}
\woctitle{New Frontiers in Physics 2014}
\newcommand{\be}{\begin{eqnarray}}
\newcommand{\ee}{\end{eqnarray}}
\newcommand{\bi}{\bibitem}

\newcommand{\rar}{\rightarrow}

\begin{document}
\selectlanguage{english}
\title{Antimatter in the universe and laboratory}
%
%

\author{A.D. Dolgov\inst{1,2,3}\fnsep\thanks{\email{dolgov@fe.infn.it
    }} 
 }

\institute{University of Ferrara, Ferrara 40100, Italy
\and
NSU, Novosibirsk, 630090, Russia
\and
 ITEP, Moscow, 117218, Russia
          }

\abstract{
Possible signatures which may indicate an existence of antimatter in the Galaxy and in the early universe
are reviewed. A model which could give rise to abundant antimatter in the Galaxy is considered. 
}
\maketitle
\section{Introduction} \label{s-intro}

86 years ago a fantastic breakthrough in particle physics was done by Paul Dirac~\cite{dirac-e+}
 who ``with the tip of his pen''  predicted a whole world of antimatter
(not just a small planet). { He assumed initially that positively charged "electron" was proton.} 
At that time physicists were rather reluctant to introduce new particles, a drastic contrast to the present
days. However, the critics by Oppenheimer that in this case hydrogen would be very unstable,
if proton was a hole in the negative continuum,
forced Dirac in 1931  to conclude that "anti-electron" is a {\it new} particle, positron, with the same mass as ${e^-}$.
Very soon after that, in 1933, Carl Anderson, discovered positron. Dirac received his Nobel prize 
immediately after this discovery, and Anderson got it three years alter, in 1936.
According to the Anderson  words: the discovery was not  was not difficult, simply nobody looked for that. 

In his Nobel lecture  ``Theory of electrons and positrons'', in December 12, 1933 Dirac said: 
``It is quite possible that...these stars being built up mainly of positrons 
and negative protons. In fact, there may be half the stars of each kind. 
The two kinds of stars would both show exactly the same spectra, 
and there would be no way of distinguishing them by present astronomical methods.''
However, we see in what follows that there are such ways and we can conclude weather a star is made of 
antimatter making astronomical observations from the Earth.

It is interesting that in 1898,
30 years before Dirac and one year after discovery of electron (J.J. Thomson, 1897),
{Arthur Schuster} (another British physicist) conjectured that there might be other sign electricity, which he
called antimatter and supposed that there might be 
entire solar systems, made of antimatter and indistinguishable from ours~\cite{schuster}.
Schuster made wild guess that  matter and antimatter are capable to annihilate and produce
energy. It happened to be ingenious and true.
He also believed that matter and antimatter  were gravitationally 
repulsive, since antimatter particles had negative mass. Two such objects on close contact
would have vanishing mass!. As we know now, this is not the case and matter and antimatter have attractive gravity.

Now we encounter at a new level these, more than century old, questions: \\
{Whether antiworlds, antistars or similar astronomically large pieces of antimatter exist in the universe?}\\
{What are observational bounds on existence of antimatter in the universe and in the Galaxy?}\\
{Do theory and observations allow for significant amount of
antimatter in our neighborhood?}

There are two roads in investigating matter-antimatter similarity and distinction:\\
{1. On the earth: theory and experiments on C, CP, and CPT breaking, e.g. 
particle and antiparticle decays or study of anti-hydrogen properties.}\\
 {2. In the sky through cosmology and astrophysics: mechanisms of baryogenesis leading to 
 antiworlds and manifestations of primordial antimatter.}
 
In what follows we discuss both ways but at different level of scrutiny.

\section{Particles and antiparticles in theories with broken CPT  \label{s-CPT}}

It is established in experiment that C, P, T, and CP symmetries are broken, while 
most probably CPT remains intact and there is a very good theoretical reason for that, namely the CPT
theorem, which states that a local, Lorenz invariant theory with canonical spin-statistics relation must be invariant
with respect to CPT-transformation~\cite{CPT}.  Nevertheless all are questioned in physics, including even the 
sacred principles and the validity of CPT theorem is being verified both in experiment and theory.  

CPT-invariance demands equality of mass of particles and antiparticles and their total decay widths, if the particle
in question are unstable:
\be 
m= \bar m\,\,\,\, {\rm and}\,\,\, \Gamma_{tot} = \bar\Gamma,
\label{m-bar-m}
\ee
but, partial decay rates of  charge conjugated channels 
and analogous cross-sections of separate charge conjugated processes may be and
are different. Total cross-section of charge conjugated reactions may also be different (due to difference of
in- and out-  multi-body states).

{Experimental tests of CPT are mostly based on measurements of mass differences of particles and 
antiparticles. In particular~\cite{pdg}:}
\be 
\frac{\Delta m_e }{m_e} = \frac{m_{e^+} - m_{e^-}} { m_{e}} < 8\times 10^{-9}.  
\label{Delta-m}
\ee
Moreover, this limit can be strongly improved because a non-zero mass difference of electron and positron
leads to breaking of gauge invariance of electrodynamics  and as a result to a non-vanishing photon 
mass~\cite{ad-van}:  
\begin{equation} 
m_\gamma^2 \sim \left(\frac{\alpha}{\pi}\right) (\Delta m)^2. 
\label{m-gamma}
\end{equation}
On the other hand, the photon mass is very strongly restricted by observations of large scale magnetic fields.
The magnetic field of the Earth gives for the Compton wave length of the photon
{${\lambda_C > 8\cdot 10^{7}}$ cm,} i.e. ${m_\gamma < 3\cdot 10^{-13}} $ eV, and
respectively {${\Delta m < 6 \cdot 10^{-12} }$~eV,} which is 
nine orders of magnitude stronger than bound (\ref{Delta-m}).

Magnetic field of the Jupiter implies
{${\lambda_C >5\cdot 10^{10}}$ cm} or ${m_\gamma <  4\cdot 10^{-16} }$ eV, and
respectively {${\Delta m < 8 \cdot 10^{-15} }$~eV.}
The strongest solar system bound comes from the analysis of the solar wind
extended to the Pluto orbit: {${\lambda_C > 2\cdot 10^{13}}$~cm.} 
 i.e. $m_\gamma <  10^{-18} $ eV. This is an
"official" limit present by the Particle Data Group~\cite{pdg}. The corresponding bound on the
electron-positron mass difference is  { ${\Delta m < 2 \cdot 10^{-17} }$~eV.} 
which is almost 14 orders of magnitude stronger than the direct bound on $\Delta m$.

The strongest of all existing bounds follows from the  observation of  large scale magnetic fields
in galaxies~\cite{Chibisov}: 
{${\lambda_C > 10^{22}}$~cm} and respectively
$m_\gamma < 2 \cdot 10^{-27} $ eV. Correspondingly 
{ ${\Delta m < 4 \cdot 10^{-26} }$ eV,} 23 orders of magnitude stronger
than the direct bound on the electron-positron mass difference.




{Similar bounds should be valid for mass differences of
any charged particle, in particular, for the mass difference of quarks.} {So small mass difference  would inhibit  baryogenesis 
in scenarios where baryogenesis is generated in thermal equilibrium because of
different masses of particles and antiparticle, see e.g.~\cite{AD-YaZ},
 but other model of CPT-odd baryogenesis due to change of kinetics, 
 induced by CPT breaking, remain feasible~\cite{ad-cpt-kin}. 

The mass differences of electrically neutral particles and antiparticles possibly is not so strongly restricted
and they may create a reasonable baryon asymmetry.

On the other hand, the mass difference is not obligatory if CPT is broken,  
there are examples of CPT-violation with ${\Delta m = 0}$ and unbroken Lorenz symmetry~\cite{cdnt}. Moreover
 CPT-odd models with ${\Delta m \neq 0}$ are rather pathological and lead, in particular, to non-conservation of 
 electric current and energy-momentum tensor~\cite{ad-vn-nonconserv}.

If CPT is broken, then there is chance to observe antistars by a difference of spectra of atoms and antiatoms created
e.g. by different masses of protons and antiprotons and electrons and positrons. However, in view of the presented
above arguments the chances of such observations are vague. Nevertheless there are effects permitting distant
observation of antistars even if CPT is an exact symmetry, as it is discussed in the next Section.

\section{Distant registration of antistars in CPT respected world \label{s-antistars} }

If CPT is unbroken, then the positions of levels in atoms and antiatoms are probably the same but
their partial decay width are surely different due to C and CP breaking. This effect was estimated
in ref.~\cite{ad-ik-ar} for hydrogen and was found to be tiny:
\be 
\Delta \Gamma /\Gamma \sim 10^{-28}.
\label{delta-Gamma}
\ee
However, an amplification, in particular, in heavier atoms and in external magnetic or electric fields might be possible

Some other ways to observe an antistar or antiplanet are considered  in ref.~\cite{ad-van-miv}.
As it is discussed in popular and/or science fiction literature, one can determine if we are dealing with matter
or antimatter through communication with inhabitants of the world under scrutiny.
It is usually assumed that to this end CP-violation can be used. If it can be established that
the light charged leptons in the shells of their atoms are more frequently produced in ${K_L}$ decays 
${K_L \to \pi^\pm e^\mp \nu}$, then we communicate with  anti-people.

Another possibility is to use polarized radio waves  asking  {whether
the polarization of charged lepton emitted in the neutron ${\beta}$ decay is the same?}
To this end CP-violation is not necessary, breaking of P would be sufficient.
However, the stellar system may be non-inhabited and even if  it is,  this process would
take an extremely long time and is more proper for a science fiction.
We need to find methods independent on intelligent life in the studied stellar system.

Neutrinos versus antineutrinos produced by thermonuclear reactions in a star might tell us if they are
emitted by matter or antimatter. However, the stellar neutrino fluxes
are too low for the present day detector sensitivity.  Neutrinos from SN explosions have better chance to be 
registered. At the first stage of SN explosion {neutrinos} from the neutronization reaction 
${pe^- \rar n\nu}$ are emitted, while antineutrinos are emitted  from anti-SN.

Detection of anti-stars by photons produced in weak interaction processes is a promising possibility.
These photons are longitudinally polarized and their energy can be well defined if they are
created in two body decays.  Mono-energetic photons produced e.g. in 
${B\rar K^\star\gamma}$ should be left-handed because of dominance of 
${b\rar s\gamma}$ penguin transition with left-handed ${s}$-quark.
{However, one can hardly imagine noticeable abundance of B-mesons in stars,}
{and most probably there is an equal amount of  ${B}$ and ${\bar B}$
mesons.}

Stars with abundant strange quarks present better chance.
The outer shell of such stars is populated by ${\Sigma}$-hyperons and 
the polarization of photons emitted in ${\Sigma^+ \rar p\gamma}$ decay could indicate
if the photons are emitted by hyperon or anti-hyperon.
The polarization is large, {${\alpha = -0.76\pm 0.08}$} and the branching ratio is nonnegligible
{${(1.23 \pm 0.05)\times 10^{-3}}$.} 

Circular polarization of photons in the ${\gamma}$-transitions of nuclei was observed
in the terrestrial experiments:
{${P_\gamma = (4\pm 1) \cdot 10^{-5}}$ in ${^{175} Lu}$ transition with the emission of 395 keV photon;} 
 $P_\gamma = -(6 \pm 1) \cdot 10^{-6}$ for 482 keV photon emitted in ${^{181}Ta}$ transition;
{${P_\gamma = (1.9 \pm 0.3) \cdot 10^{-5}}$
for 1290 keV photon emitted in transition of ${^{41}K}$.}
Measurement of circular polarization of such photon lines is ideally suited for search of antistars.

\section{Search for cosmic antinuclei \label{s-anti-nucl}}

An unambiguous indication of cosmic antimatter could be registration of helium, or heavier, antinuclei 
in cosmic rays. Their secondary production is strongly suppressed. Thus a registration of e.g. $^4 \bar He$
at the level exceeding theoretical expectations for secondary production (see below) would prove that
antimatter objects are indeed present in the universe. 

There are three detectors for search of antinuclei in cosmic rays:\\
1. BESS: Japanese Balloon Borne Experiment with Superconducting
Solenoidal Spectrometer. At the present time they have the most stringent upper
bound on the ratio of fluxes: $ \bar He / He < 10^{-7}$~\cite{BESS-He}. \\
2. PAMELA (Italian-Russian space mission): Payload for Antimatter 
Matter Exploration and Light-nuclei Astrophysics. Their bound is somewhat less restrictive: 
$ \bar He / He < 4.7\cdot 10^{-7}$~\cite{PAMELA-He} . Presumably it may be improved by an order of magnitude.\\
3. AMS: AntiMatter Spectrometer (Alpha Magnetic Spectrometer). Presently they have the 
limit~$ \bar He / He <  10^{-6}$ based on AMS-1 data~\cite{AMS-He}, which may possibly be improved by three orders of 
magnitude on AMS-2.

The observed flux of cosmic helium at  ${ E< 10}$ GeV/nuclei is 
$ dN/dE = 10^{2}$ /m$^2$/str/sec/GeV.
The expected fluxes of secondary produced anti-nuclei are the following.
According to the calculations of ref.~\cite{antinucl} deuterium produced in  ${\bar p\, p}$
or ${\bar p\, He}$ collisions should have the flux$\sim 10^{-7} $/m$^{2}$/ s$^{-1}$ /sr/(GeV/n),
i.e. 5 orders of magnitude lower than the observed flux of antiprotons.
{The expected fluxes of secondary produced 
${^3\bar He}$ and ${^4\bar He}$ 
are respectively 4 and
8 orders of magnitude smaller than the flux of anti-D.}

So to summarize, the  observations and bounds are the following:\\
{${ \bar p / p \sim 10^{-5}-10^{-4}}$, observed, can be explained by secondary production;}
${He/p \sim 0.1}$.
{The upper limit: ${\bar He / He < 3\times 10^{-7}}$}.
Theoretical predictions: ${ \bar d \sim 10^{-5} \bar p}$,
${ ^3\bar He  \sim 10^{-9} \bar p}$,
${ ^4\bar He \sim 10^{-13} \bar p}$.
{From the upper limit on ${\bar{He}}$: the nearest
single antigalaxy should be further than} 
{10 Mpc (very crudely).}

According to the data of ALICE detector~\cite{alice} obtained at LHC, production of an antinucleous with an
additional antinucleon is suppressed only by factor about 1/300 which is much milder than the suppression
factors presented above. Probably the difference is related to much higher energies at which data of ALICE are
taken. The events with such energies are quite rare in cosmic rays.

\section{Bounds on cosmic antimatter based on indirect data \label{indirect}}

A signal of antimatter may be delivered by an intense gamma ray line with
0.511 MeV energy, which surely comes from $e^+ e^-$ - annihilation at rest.  Such a line is
observed coming from the Galactic center, see e.g. ref.~\cite{positron-line}. 
A little later SPI spectrometer has also provided evidence for the disk or halo 
component~\cite{spi-halo} of this line. 
However, it is not excluded that the source of this line could be secondary positrons. 

Proton-antiproton annihilation creates a wide spectrum of gamma rays with the energies about
 100 MeV. If there is an antimatter galaxy in our cluster, then the annihilation of infalling intergalactic 
gas with antimatter in the antigalaxy would produce a constant flux of $\sim 100$ MeV photons which would
be well observed on the Earth if the distance to such antigalaxy is shorter than 10 Mpc~\cite{steigman-76}.
Similar considerations allow to restrict the fraction of antimatter in two colliding galaxies in
Bullet cluster down to $10^{-6}$~\cite{steigman-2008}.

If there is an antistar in the Galaxy, then the Bondi accretion of interstellar gas to its surface 
would result in the gamma-ray luminosity~\cite{ballmoos}:
\be 
 L_\gamma \sim 3\cdot 10^{35} (M/M_{\odot})^2 v_6^{-3}\,\,{\rm erg/sec},
\label{L-gamma}
\ee
where $v_6$ is the star velocity with respect to gas in $10^6$ cm/sec units. Typically $v_6 \sim 1$.
An absence of such bright gamma-ray sources put  the following limit on a possible number of antistars in  
vicinity of the Earth {${N_{\bar *} / N_{*} <  4\cdot 10^{-5} }$} inside 150 pc from the Sun. 

Quite strong bound on existence of astronomically large antimatter domains was derived in ref.~\cite{CdiRG}, namely
that in the charge symmetric universe the nearest antimatter domain must be at least at the distance of a few Gpc.
 
A set of bounds can be derived from the condition that the observed angular fluctuations of CMB exclude large
isocurvature fluctuations at distances larger than, roughly speaking, 10 Mpc. Big bang nucleosynthesis does not
allow noticeable chemistry fluctuations at  $ d > 1$ Mpc. This is the scale which is of the order of the comoving galactic
scale and below it inhomogeneous chemical content would be homogenized because of mixing of matter in galaxies.

However, the presented bounds are true if antimatter makes the 
same type objects as the observed matter.
{For example, compact faster objects made of antimatter may be abundant
in the Galaxy but still escape observations, as it is discussed below. In particular, the antimatter objects, which we consider here, 
are supposed to be created in the very early universe and behave as dark matter, mostly populating the galactic haloes. So their
velocity should be of the order of a few hundred kilometers per second and their luminosity (\ref{L-gamma}) due to the annihilation on the
surface would be strongly reduced.

\section{C and CP violation in cosmology \label{s-CP-cosm}}

Breaking of C and CP invariance is established in experiment and very naturally appears in the (minimal)
standard model (MSM) of particle physics. However, the mechanisms of C- and CP-violation realized in
cosmology may have nothing to do with those which could be realized in particle physics. At first sight
an existence of additional mechanisms of charge symmetry breaking is at odds with the Occam's razor: 
"entities must not be multiplied beyond necessity" (``entia non sunt multiplicanda praeter necessitatem''). However,
other ways to break C or/and CP in cosmology are so natural that they simply must exist. More detail about
charge symmetry breaking can be found in lectures~\cite{ad-30,ad-varenna}.

{There are three possibilities to break CP in cosmology:} \\
{1. Explicit}, the usual one operating in particle physics by allowing complex constants in the Lagrangian. Normally leads
to the universe without antimatter. \\
{2. Spontaneous,} locally indistinguishable from the explicit one, but leading
to globally charge symmetric universe.\\
{3. Stochastic or dynamical,} unobservable in particle physics. May lead to noticeable amount of 
antimatter and globally asymmetric universe.

I. Explicit CP- breaking is induced if some coupling constants or masses in the Lagrangian have nonzero
imaginary part. For example it is realized in MSM if the non-diagonal coupling constant of the Higgs boson to
different quarks is complex, $H ( g \bar \psi_i \psi_j + g^* \bar \psi_j \psi_i)$ with $i\neq j$. Vacuum condesate of the 
Higgs field leads to a complex mass matrix of three quark families where the non-zero phase cannot be rotated away.
This is a very natural mechanism of CP violation in particle physics, which well agrees with experiment. However, this
mechanism is too weak, roughly speaking by 10 orders of magnitude, to explain the observed baryon asymmetry.

II. Spontaneous CP-violation~\cite{lee-cp} is analogous to spontaneous symmetry breaking in gauge theories.
The Lagrangian of the theory is supposed to be C and/or CP invariant but the vacuum state is not. The vacuum is
usually doubly degenerate and a complex scalar field, ${\phi}$,  acquires different vacuum expectation values
over these two vacua:
\be
\langle \Phi \rangle = \pm f\,,
\label{Phi-spont}
\ee
so the signs of C/CP - violation over different vacuum states are opposite. Locally in the laboratory this mechanism
indistinguishable from the explicit one. However, if such mechanism were realized in
cosmology the universe would globally be charge symmetric (50:50 matter and antimatter) consisting of several (many?) 
matter and antimatter  domains. 

This mechanism encounters serious cosmological problems. Firstly, the walls separating different domains
have huge surface energy density which would destroy the isotropy of the universe~\cite{zko}.
To avoid that the nearest domain of antimatter should be far beyond cosmological horizon
${ l_B \gg }$ Gpc} or a mechanism of wall destruction should operate.

The size of domains are normally too small and matter and antimatter
would annihilate leaving behind an empty universe. This problem could be 
solved with mild inflationary expansion after CP-breaking took place~\cite{sato}. 
As we mentioned in the previous section, the observational bound on the size of the domains which appeared as a result
of spontaneous CP-violation, derived from the data on the gamma-ray background, is about a few Gpc~\cite{CdiRG}.

III. Dynamical or stochastic CP-violation~\cite{ad-varenna} could be achieved by a complex scalar field displaced from 
the equilibrium point. It may happen e.g. at inflationary stage due to rising quantum fluctuations of light fields. 
If this field relaxed to equilibrium after baryogenesis had been done, it can supply sufficient C/CP violation for successful
generation of the cosmological charge asymmetry. Such mechanism of CP-violation would operate only in the early
universe and disappear without trace today. Of course the domain walls  disappeared together with the field and
do not create any problem.

So, why do we multiply entities against Occam?  
Nature is usually very economical and usually do not make unnecessary efforts. 
However, this  mechanism of CP violation is quite natural because it always operated 
in the early universe if there exists a complex scalar field with ${m<H_{inf}}$.}
Such temporary CP-violation could give rise to an 
inhomogeneous baryon asymmetry, ${ \beta(x)}$, with possible antimatter nearby.

\section{Baryogenesis \label{s-bg}}

The prevailing point of view at the present time is that all the universe is made only of matter and there is
no primeval antimatter. Still the fact that antimatter exists 
created fundamental cosmological puzzle: why the observed universe is 100\% dominated by matter?
Antimatter exists but not antiworlds, why? The problem deepened because of approximate 
symmetry between particles and antiparticles.

The puzzle of the observed predominance of matter over antimatter 
was resolved by Sakharov~\cite{sakharov} on the basis of the conditions:\\
{I. Nonconservation of baryons.} \\ {
II. Violation of symmetry between particles and antiparticles,  i.e. C and CP breaking.} \\
III. Deviation from thermal equilibrium.\\
However, none of these three conditions is obligatory. For a review of different scenarios of baryogenesis, 
where some of these conditions are not fulfilled, see e.g. ref.~\cite{ad-30}

{There is plethora of baryogenesis scenarios for explanation of just one number:}
\be 
\beta_{observed} = \frac{N_B-N_{\bar B}}{N_\gamma} 
\approx 6\times 10^{-10}\,.
\label{beta-obs}
\ee
The usual outcome of all these scenarios is ${ \beta = const}$, 
which makes it impossible to distinguish between models and
does not leave space for cosmological antimatter.
All the models, but one, by Affleck and Dine (AD)~\cite{ad-bs} give rise to a small
${ \beta}$ but {this particular model may create ${\beta \sim 1}$.}
  
Natural generalization of AD scenario allows for a lot of antimatter almost at hand, i.e.  plenty of antimatter
objects in the Galaxy. An observation of cosmic antimatter will give a clue
to baryogenesis, to the mechanism of cosmological
C and CP breaking, and present an extra argument in  favor of inflation.
Since generalized scenarios predict a whole function {${\beta (x)}$,} the models are falsifiable.

In what follows we consider a modification~\cite{ad-js}  of the AD scenario and show how this modification can lead 
to abundant antimatter objects which are still escaping observations. As is known supersymmetric theory predicts 
an existence of scalar partner of baryon which has nonzero baryonic number, ${ B_\chi\neq 0}$. 
Such scalar bosons may condense along {flat} directions of the potential, which are generally present.
In a toy model the potential of $\chi$ has the form:
\be
U_\lambda(\chi) = \lambda |\chi|^4 \left( 1- \cos 4\theta \right),
\label{U-of-chi}
\ee
where  field $\chi$ is taken in the form: ${ \chi = |\chi| \exp (i\theta)}$.
In GUT SUSY theories  baryonic number is naturally non-conserved, and this is reflected in the absence of the symmetry
with respect to the phase rotation of $\chi$ in this potential. Due to rising quantum fluctuations $\chi$  could "travel" away from
zero along flat directions of the potential: $\cos 4\theta =1$. This happened during inflation, when the Hubble parameter was
larger than the mass of $\chi$. When inflation was over and $H$ dropped below $m_\chi$, the field started to move down to the
origin, according to the equation of motion:
\be
\ddot \chi +3H\dot \chi +U' (\chi) = 0.
\label{ddot-chi}
\ee

If it moved strictly along the bottom of the valley, it would come to the origin with zero angular momentum $B_\chi = \dot \theta |\chi|^2$,
which is the baryonic charge of $\chi$. However, if there are other flat directions, which are important at small $\chi$, the field moves from
one valley to another, acquiring nonzero angular momentum, i.e. nonzero baryonic number density. It would happen if the potential
contains the mass term, which leads to:
\be
U_m( \chi )  = { m^2 \chi^2 + m^{*\,2}\chi^{*\,2}}= m^2 |\chi|^2} {\left[{ 1-\cos (2\theta+2\alpha)}  \right],
\label{U-m-chi}
\ee
where ${ m=|m|e^\alpha}$. {If ${\alpha \neq 0}$, C and CP are explicitly broken.}

When ${{\chi}}$ decays into quarks through B-conserving process, its baryonic charge is transferred to the baryon asymmetry of quarks.

The modification of the usual AD scenario suggested in ref~\cite{ad-js} consists in the addition of interaction of $\chi$ with the inflaton
field $\Phi$ of general renormalizable form:
\be 
U = g|\chi|^2 (\Phi -\Phi_1)^2 , 
\label{U-pf-chi-phi}
\ee
where $\Phi_1$ is the value of the inflaton field which it passed during inflation not too long before it was over. This is the 
only tuning of the model.

This interaction gives rise to time dependent effective mass squared of $\chi$, proportional to the difference $(\Phi-\Phi_1)^2$.
So the window to flat direction is open only during a short period, when $\Phi$ is close to $\Phi_1$. Hence with a small
probability $\chi$ could reach a large value. So
cosmologically small but possibly astronomically large  bubbles with high ${ \beta}$ could be
created, occupying a small fraction of the universe, while the rest of the universe has normal
{${{ \beta \approx 6\cdot 10^{-10}}}$, created by small ${\chi}$}. In this way bubbles with high baryonic number density
would be created. Since the direction of $\chi$ rotation, i.e. the sign of its
baryonic number, is chaotic in the simplest version of the model, there would be an equal amount of baryonic and
antibaryonic bubbles. So the model predicts that a large amount  of baryons and probably an equal amount of antibaryons
should be in the form of compact stellar-like objects or primordial black holes (PBH). The rest of baryons (and no antibaryons) 
are in the form of the observed smooth baryonic background, created by the same baryogenesis mechanism with small initial $\chi$.
The amount of antimatter may be comparable or even larger 
than that of known baryons, but such ``compact'' (anti)baryonic objects
would not contradict  BBN and CMB and even make (some?) dark matter.

\begin{figure}[h]
\centering
\includegraphics[width=7cm,clip]{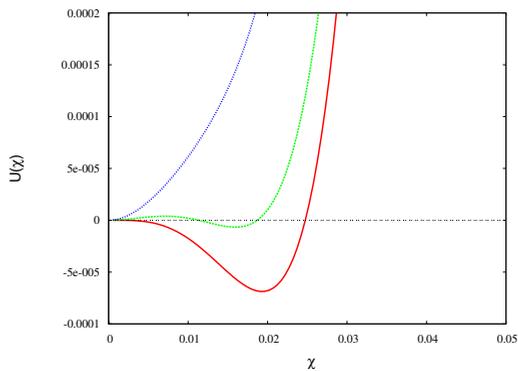}
\caption{Evolution of the potential of $ \chi$ as a function of the
inflaton field $ \Phi$. The potential has initially  the only minimum at $\chi=0$. When $\Phi$ approaches $\Phi_1$ the minimum shifts to 
$\chi \neq 0$. Next with rising $\Phi$ the initial picture is restored. }
\label{fig-1}       
\end{figure}

The distributions of high baryon density bubbles over length and mass have the log-normal form:
\be
\frac{dN}{dM} = C_M \exp{[-\gamma \ln^2 (M/M_0)]},
\label{dn-dM}
\ee
where ${C_M}$, ${\gamma}$, and ${M_0}$ are some constant model dependent parameters.
The spectrum is practically model independent, it is determined by the field evolution during inflation. 

The density contrast between regions with high and low baryonic number density initially was practically
zero till the QCD phase transition  at ${\ T\sim 100}$ MeV when quarks confined and formed non-relativistic protons.
In this process PBH with masses from a few solar masses ${M_\odot}$, to ${10^{6-7} M_\odot}$ could be created.
This mechanism explains an early formation of quasars, which remains mysterious otherwise. It can also explain an 
existence of stars "older than the universe" and supernovae and gamma-bursters at high redshifts~\cite{ad-sb}.
Since "our" black holes were formed in the very early universe at $t\sim 100$ sec, they might be seeds for subsequent
galaxy formation. Smaller mass compact baryo-dense stars may be the observed Machos. Anyhow
 the universe may be full of early formed and by now dead stars and antistars.

Another interesting feature of the considered model is a prediction of anomalously high production of metals at big bang
nucleosynthesis~\cite{bbn-anom}.
 It may explain evolved chemistry observed around high redshift quasars. Moreover, an existence of
(compact) stars with unusual chemistry is predicted. So if such a star is observed, it could consist of antimatter with 50\%
probability. Thus if a cloud or compact object  with anomalous chemistry is found, search for matter-antimatter  annihilation there.

Observational manifestations of such early formed compact stellar-like matter and antimatter objects are studied in
refs.~\cite{bambi-ad,sb-ad-kp}.

\section{Cosclusion \label{s-concl}}

1. The Galaxy may possess a noticeable amount of antimatter which still escaped observation. \\
2. Theoretical predictions for the amount of cosmological  antimatter are vague and model dependent, but testable
and may permit to distinguish between different  scenarios of baryogenesis.\\
3. Not only ${ ^4 \bar{He}}$ is worth to look for but 
also heavier anti-elements. Their abundances could be similar to those observed in
SN explosions.\\
{4. Regions with an anomalous abundances of light elements are suspicious
that there may be anti-elements.}\\
{5. A search of cosmic antimatter has nonvanishing chance to be 
successful.} \\
6. Dark matter made of BH, anti-BH, and dead stars is a promising
candidate. There is a chance to understand why ${ \Omega_B =0.05}$ is 
similar to ${ \Omega_{DM} = 0.25}$.\\
{7. Detection of $ { \bar \nu}$ in the first burst from anti-SN
explosion would allow to check if a star or antistar was exploded.\\  
8. Measurement of polarization of different forms of  electromagnetic radiation coming from stars would indicate if it is 
a star or an antistar. \\
{9. Signatures in favor of cosmic antimatter:}\\
{The observed 0.511 MeV line from the galactic bulge
and especially (if confirmed) from the halo can be a signature of cosmic antimatter!}\\
{Unidentified EGRET sources may be antistars.} \\
Less direct possible evidence: supermassive black holes at high redshifts; enriched with metals gas around early quasars; stars, older that the universe...
\\[4mm]
{\bf Acknowledgement} This work was  supported by the grant of the Russian Federation government
11.G34.31.0047.

\end{document}